\documentclass{article}
\usepackage[square, sort, numbers]{natbib}
\usepackage[final]{nips_2017}
\usepackage[utf8]{inputenc}
\usepackage[T1]{fontenc}
\usepackage{textcomp}
\usepackage{gensymb}
\usepackage{hyperref}       
\usepackage{url}            
\usepackage{booktabs}       
\usepackage{amsfonts}       
\usepackage{nicefrac}       
\usepackage{microtype}      
\usepackage{graphicx}
\usepackage[english]{babel}
\usepackage{tabularx}
\usepackage{amsmath}
\usepackage{wrapfig}
\usepackage{ amssymb }
\usepackage{float}
\usepackage{subfigure}
\usepackage{caption}
\usepackage{xcolor}
\usepackage{mhchem}

\definecolor{cardinal}{HTML}{8c1515}

\title{Modeling Sensorimotor Coordination as Multi-Agent Reinforcement Learning with Differentiable Communication}

\author{
  Bowen Jing \\
  Department of Computer Science \\
  Stanford University\\
  \texttt{bjing@stanford.edu}  \\
  \And
  William Yin \\
  Department of Symbolic Systems \\
  Stanford University \\
  \texttt{wyin@stanford.edu} \\
}

\begin{document}

\begin{center}
\end{center}

\maketitle
\begin{abstract}
Multi-agent reinforcement learning has shown promise on a variety of cooperative tasks as a consequence of recent developments in differentiable inter-agent communication. However, most architectures are limited to pools of homogeneous agents, limiting their applicability. Here we propose a modular framework for learning complex tasks in which a traditional monolithic agent is framed as a collection of cooperating heterogeneous agents. We apply this approach to model sensorimotor coordination in the neocortex as a multi-agent reinforcement learning problem. Our results demonstrate proof-of-concept of the proposed architecture and open new avenues for learning complex tasks and for understanding functional localization in the brain and future intelligent systems.
\end{abstract}

\section{Introduction}	

Motor coordination tasks have been among the most popular applications for reinforcement learning models and significant milestones have been achieved in a diverse array of tasks  in environments like MuJoCo\cite{MuJoCo} and OpenAI Gym\cite{OpenAI}. Nearly all models of motor coordination have followed the paradigm of a single agent learning to process observations from the environment and direct all actions accordingly. In this paradigm, a single agent is analogous to a single individual who learns to perform a certain motor task.

In this paper, we propose modeling the sensorimotor cortex as a \textit{collection} of individual agents. We draw major inspiration from the neuro-anatomical localization of visual processing, proprio-sensory perception and motor planning to the occipital, parietal, and frontal lobes, respectively. Due to the spatial separation of these functions in the brain, we explore the perspective that each cortex performs internal computations on significantly shorter timescales and signicantly higher bandwidths than messages between cortices. Given this, it is more appropriate to model the sensorimotor cortex as a group of smaller agents which communicate and coordinate their actions to achieve a broader goal. Our approach is made possible by recent developments in multi-agent reinforcement learning which permit the learning of differentiable communication protocols amongst cooperating agents, and to our knowledge is novel in the literature.

\section{Multi-Agent Reinforcement Learning}

Multi-agent reinforcement learning (MARL) is a promising approach for modeling the dynamics of cooperative, competitive, or predator-and-prey relationships between agents, where each agent has some unique observation of the state of the world and then proceeds to make individual decisions which contribute to the world state at the next timestep. A recent example from Liu \textit{et al.} — in which emergent cooperative behaviors were studied in agents playing a game of soccer modeled in MuJoCo — highlights the extraordinary potential for multi-agent learning within such a model\cite{2019arXiv190207151L}, as does the work of Bard \textit{et al.} on Hanabi, a cooperative card game\cite{2019arXiv190200506B}.

The mode of communication between agents remains an active area of research. Foerster \textit{et al.}\cite{2016arXiv160506676F} recently proposed two means of inter-agent communication: Reinforced Inter-Agent Learning (RIAL) and Differentiable Inter-Agent Learning (DIAL). RIAL describes an approach which employs deep Q-learning and allows communication during action selection, while DIAL describes a means to backpropogate error derivatives through noisy communication channels. DIAL is particularly interesting because it employs centralized learning but decentralized execution; in other words, learning occurs across all agents as one unit, while action selection happens on a per-agent basis where each agent treats the other agents as elements of the environment. On the other hand, Sukhbaatar \textit{et al.}\cite{2016arXiv160507736S} describe the development of a neural model as a continuous communication channel between agents (called CommNet) which facilitates the learning of a communication protocol among agents. In both of these examples, the protocol itself is not specified to the agents; rather, the agents are 'tasked' with learning the protocol themselves, and in doing so revealed interesting strategies to approaching inter-agent communication and information transfer.

Despite these advancements, however, state-of-the-art MARL models still remain relatively structurally simple. In particular, agents in MARL models remain relatively homogeneous, in that agent parameters are drawn from the same distributions. The agents themselves also have the same objectives, regardless of whether the task is cooperative or competitive. In this sense, these MARL environments frequently fail to extend well to scenarios in which different agents have different roles or objectives. 

In this sense, we see extraordinary potential in extending MARL models to scenarios in which agents may be defined with unique roles and properties. In particular, we hypothesized that such a model would be especially appropriate in modeling motor control tasks. There already exists a wealth of research in employing policy gradient techniques in modeling motor skills and control\cite{Peters}, but none which aim to model each component of a motor agent as a unique agent with unique properties. The sensorimotor cortex seems to be a perfect candidate for such a model due to the separation of distinct components which interact\cite{separation}. Hence, our paper proposes the use of continuous communication protocols for message-passing across distinct, role-specific sensorimotor agents as models for components of a motor control task. 

\section{Architecture}
\subsection{Model}
In our architecture we model the sensorimotor cortex as a set $\mathcal{S}$ of $n_s$ sensory agents and a set $\mathcal{M}$ of $n_m$ motor agents embedded in a fully deterministic Markov decision process with a set of discrete or continuous observation spaces and a set of discrete action spaces\footnote{The model can easily be adapted to continuous action spaces.}. Each sensory agent $S_i \in \mathcal{S}$ makes observations $o^{[t]}_i$ in some input domain and outputs a message $m_{i,j}$ to each motor agent\footnote{Each motor agent is an ordered pair, as explained in the following paragraph.} $(M_j, Q_j) \in \mathcal{M}$. The sensory agents do not have access to the decision process and do not themselves take any actions. Rather, their sole effect is achieved by communicating a summary of the sensory information to the motor agents.

Each motor agent consists of an M-net $M_i$ and a Q-net $Q_i$. The M-net takes in is all incoming messages (from all sensory agents as well as all other M-nets), in addition to the most recent action $a^{[t-1]}_i$ of the corresponding motor agent, and produces a message for each of the other motor units $(M_j, Q_j) \in \mathcal{M}, i\neq j$. The Q-net is a standard action-value function which takes as input all incoming messages from sensory agents and all motor units, as well the previous action $a^{[t-1]}_i$ and predicts a value for each possible action $a_i$ given the input messages. Messages are not sent from the M-net of each motor unit to its corresponding Q-net because doing so would give encourage the M-net to serve as intermediate information-processing agents unaffiliated with a particular Q-net.

In our model of message passing, we set that each message is received by the receiving agent one time step after it is sent. This serves dual purposes: first, it captures the longer period of time it would take information to propogate from one part of the cortex to another; second, it enables the training of a message protocol via gradient descent despite a cyclic communication network.

More formally, each of our model's sensory agents $S_i \in \mathcal{S}$ computes:
\begin{equation*}
    \left(m^{[t]}_{i,1} \dots m^{[t]}_{i,n_m}\right) = S_i\left(o^{[t]}_i\right)
\end{equation*}
where $m_{i,j}$ indicates a message vector from sensory agent $S_i$ to motor agent $M_j$. Each of the model's motor units $(M_i, Q_i) \in \mathcal{M}$ computes
\begin{eqnarray*}
\left(\dots m'^{[t]}_{i,i-1}, m'^{[t]}_{i,i+1} \dots\right) &=&  M_i\left(m^{[t-1]}_{1,i} \dots m^{[t-1]}_{n_s,i}, \dots m'^{[t-1]}_{i,i-1}, m'^{[t-1]}_{i,i+1}\dots, a^{[t-1]}_i\right) \\
Q_i^{[t]}(a) &=& Q_i\left(m^{[t-1]}_{1,i} \dots m^{[t-1]}_{n_s,i}, \dots m'^{[t-1]}_{i,i-1}, m'^{[t-1]}_{i,i+1}\dots, a^{[t-1]}_i, a\right)
\end{eqnarray*}
where 
\begin{itemize}
    \item $m'_{i,j}$ indicates a message vector from motor agent $(M_i, Q_i)$ to motor agent $(M_j, Q_j)$
    \item $\dots m'_{i,i-1}, m'_{i,i+1} \dots$ denotes the set of all $m'_{i,j}$ where $i\neq j$
    \item $a$ is an element of $A_i$, the action space of agent $(M_i, Q_i)$
    \item $Q^{[t]}_i$ is the action-value function for motor agent $(M_i, Q_i)$ at time $t$
\end{itemize}    
for a total of $n_m(n_m + n_s-1)$ message streams. The controller then chooses the actions for each motor agent according to
\begin{equation*}
    a^{[t]}_i \sim \text{soft}\max_{a \in A_i} Q^{[t]}_i(a)
\end{equation*}
if the controller is exploratory / learning mode and 
\begin{equation*}
    a^{[t]}_i = \arg\max_{a \in A_i} Q^{[t]}_i(a)
\end{equation*}
if it is in greedy mode. At each timestep $t$, the environment provides a reward $R^{[t]}$ in response to the actions $a_i^{[t]}$ taken. We train the action-value functions to predict $Q^{[t]}_i(a_i^{[t]}) =\mathbb{E}[\sum_{\tau = t}^\infty \gamma^{\tau-t}R^{[\tau]}]$ by performing stochastic gradient descent on the temporal-difference loss
\begin{equation*}
    \left(Q^{[t]}_i(a_i^{[t]}) - R_i^{[t]} - \gamma \arg\max_{a\in A_i} Q^{[t+1]}_i(a)\right)^2
\end{equation*}.
\subsection{Implementation}
For our training environment, we instantiate the architecture to feature three sensory agents and two motor agents. The sensory agents $S_l$, $S_e$, and $S_r$ represent the left arm propriosensory cortex, the visual cortex, and the right arm propriosensory cortex, respectively, and the motor motor agents $(M_l, Q_l)$ and $(M_r, Q_r)$ controls the left and right arm, respectively. The observation domains $S_l, S_r \in \mathbb{R}^2$ represent the $(x,y)$ coordinates of the respective arm (we restrict the environment to two dimensions for simplicity), while $S_e \in \mathbb{R}^4$ encodes the $(x, y)$ location and $\Dot{x}, \Dot{y}$ velocity of an object of interest --- in our case, a ball\footnote{In a more sophisticated system, the sensory agent may instead be a convolutional network operating on an input image stream.}. The action spaces $A_l$ and $A_r$ consist of five actions: moving the respective arms a fixed amount in one of four directions, or making no movements\footnote{In an ideal, more realistic model of muscle movement, the outputs would affect the acceleration of the arm.}.

We implement each functional component of the system (the sensory agents, the M-nets, and the Q-nets) as a neural network with one ten-neuron hidden layer and ReLU activations operating on a concatenation of its inputs and outputting a concatenation of its outputs. While the architecture permits each message stream to be of different length, we implement all message channels to be of fixed dimensionality $d_m$ = 5. This fixed dimensionality, the number of layers and hidden units in each neural network, the learning rate, and the discount factor $0 < \gamma < 1$ are the hyperparameters of our model.

The model is trained by regarding the $o^{[t-1]}$s, $m^{[t-2]}$s, and $m'^{[t-2]}$s as constant inputs to the model whose output is the set of $Q^{[t]}$ functions at time $t$. Gradient descent is then simultaneously performed on all the network weights.
\section{Environment}

We train the model on a simplified juggling-like task, in which the model must bounce a ball using paddles in each hand and prevent it from touching the ground (Figure 1). Collisions are perfectly elastic, such that the total number of bounces is principally unbounded and the bounce frequency and heights is relatively constant\footnote{This has the same effect as the more realistic assumption that the agent imparts an compensatory upward impulse after a non-zero contact time, but that more sophisticated learning requirement may be better suited for a larger model.}. Each bounce imparts a Gaussian-distributed random horizontal impulse to the ball; if the ball hits the wall, it is deflected. The simulation resets when the ball hits the ground. Rewards are as follows:
\begin{itemize}
    \item A penalty of 50 if the ball hits the ground and the simulation resets.
    \item A contact award of 50 each time the agent successfully bounces the ball. We found this shortening of the time horizon to be necessary for a resonable rate of learning.
    \item To incentivize coordination between the motor agents, a global penalty of 0.5 is applied to each movement of any arm\footnote{An alternative perspective is to apply a penalty whenever both agents move.}.
\begin{figure}[H]
    \centering
    \includegraphics[scale=0.48]{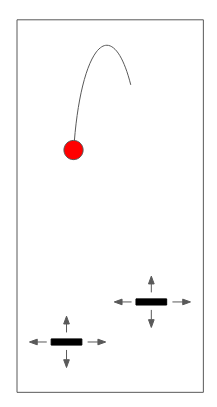}
    \captionsetup{justification=centering}
    \caption{The environment consists of a ball which the agent must keep in the air by moving paddles in each hand. See details in text.}
    \label{fig:my_label}
\end{figure}
\end{itemize}
The values of these rewards, the bounce height, and simulation frequency may be adjusted as hyperparameters. In our implementation of the environment, the physical parameters of the model are analogous to a human agent with arms at height 1m, bouncing a ball dropped from a height of 3m, with paddles 30cm across, in a bounding box of width 2m, arm movements of 15cm, and a message transit time of 0.1 seconds\footnote{Of course many alternative formulations exist, but we choose this distance scale for intuitive simplicity.}.

\section{Experiments}
We train each of $n=10$ instances of the model for 5000 sessions, where each session is defined as ending when the ball hits the ground. We alternate 100-session exploratory learning epochs with 100-session greedy learning epochs\footnote{The exploratory and greedy controllers are as described above. Alternatively, the "temperature" of the softmax is may be tuned.}. The greedy epochs revealed that the model learned to minimize movements by restricting action to one of the two agents while the other agent became increasingly dormant (Figure 2). This was an interesting and unexpected result, as it is highly analogous to the notion of hand dominance. However, no such trend was observed in the exploratory epochs, suggesting the preference was very slight (Figure 3). Taken at face value, the hand dominance in our model developed because the model found it more effective to universally suppress the action of one agent rather than trying to develop nuanced communications among agents.

\begin{figure}[H]
    \centering
    \includegraphics[scale=0.6]{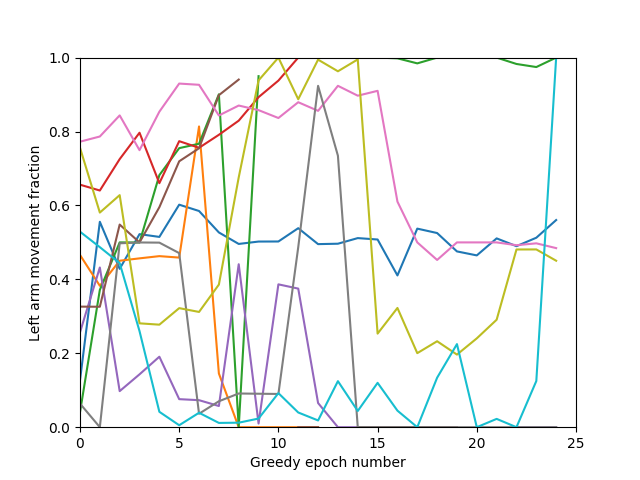}
    \captionsetup{justification=centering}
    \caption{Left arm movements as a fraction of total movements during the greedy training epochs. Notice that by the later epochs, a majority of models have a "dominant" or preferred agent to take action with.}
    \label{fig:my_label}
\end{figure}

\begin{figure}[H]
    \centering
    \includegraphics[scale=0.6]{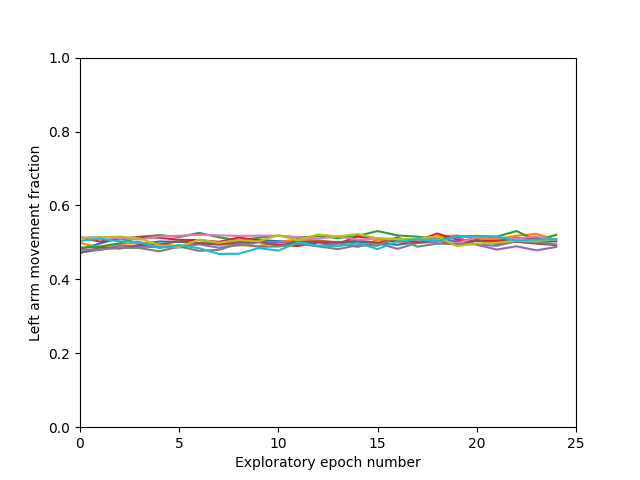}
    \captionsetup{justification=centering}
    \caption{Left arm movements as a fraction of total movements during the exploratory training epochs. No significant deviation from an even distribution is observed.}
    \label{fig:my_label}
\end{figure}

Figures 4 and 5 show the performance of the multi-agent model over the course of training. Despite performing the best among reasonable configurations explored\footnote{We were able to get much more impressive results by making the paddle comically large and the learning task almost trivial.}, the model appears to exhibit only relatively weak learning. As shown in Figure 4, the mean number of bounces per 100-session greedy learning epoch rose from 26.5 to 41.8 in the first 10 epochs ($t(18)=-2.816, p=0.011$) but no significant further improvement was observed in the remaining 15 epochs ($t(18)=-0.327, p=0.747$). Note that this corresponds to an average of only 0.4 bounces before the ball hits the ground, even at the end of training. In the exploratory learning epochs, performance remained at around 37 bounces per epoch throughout, with no signs of learning observed (Figure 5). This is reasonable, as the model may not be confident enough yet in the learned policy to achieve consistent performance by drawing actions from a softmax rather than an argmax.

\begin{figure}[H]
    \centering
    \includegraphics[scale=0.6]{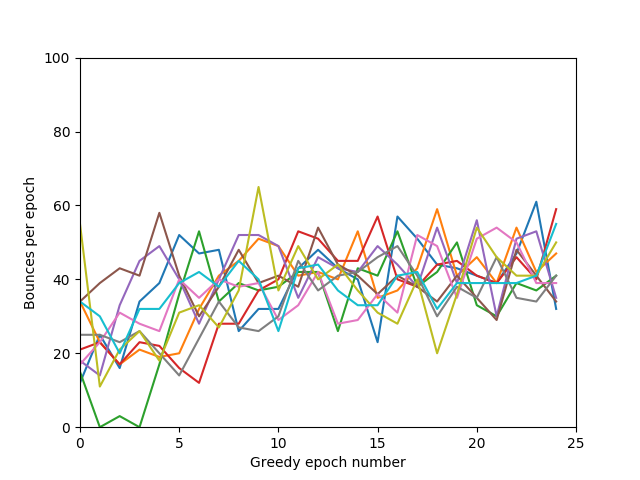}
    \captionsetup{justification=centering}
    \caption{Number of bounces per greedy learning epoch.}
    \label{fig:my_label}
\end{figure}

\begin{figure}[H]
    \centering
    \includegraphics[scale=0.6]{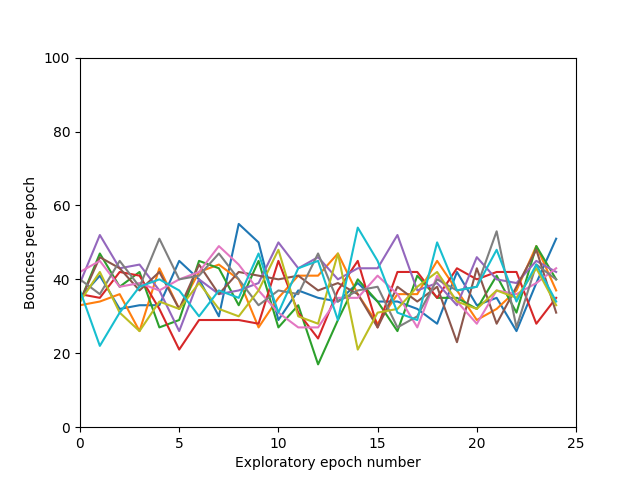}
    \captionsetup{justification=centering}
    \caption{Number of bounces per exploratory learning epoch.}
    \label{fig:my_label}
\end{figure}

\section{Discussion and Future Work}

We have demonstrated the implementation of a multi-agent model of the sensorimotor cortex trained on a simple ball-bouncing task. Although further hyperparameter tuning may improve performance to reasonable human-level expectations, we have demonstrated proof of principle that the proposed architecture is capable of learning a sensorimotor coordination task. In particular, we have demonstrated that delayed message passing from \textit{sensory} agents to \textit{motor} agents is sufficient to allow the learning of sensory-guided motor actions.

Further work is merited to confirm the model of message-passing between motor agents as a means of coordinating actions. Such validation could be performed by ablation studies in which message-passing between such agents is removed. The hand-dominance observed in the present experiments offers no evidence of multi-agent coordination, but is instead the chief unexpected result of this paper. Because all awards and penalties are shared by both agents, there is no feature of the architecture which would encourage the unilateral, consistent suppression of a single agent. It is possible that the stochastic preferential usage of a particular agent early in training reinforces differences in the learned Q-values in such a way that further biases the model towards the usage of that agent. The trajectories in Figure 2 support this possibility, as they generally converge toward the dominance which was present to a slight degree in early epochs. If so, then hand dominance would appear to be a stable configuration which naturally arises from an motion-conservative reward system and trial-and-error learning. It is worth investigating or speculating if this has any connection with the formation of hand-dominance in humans.

\small

\bibliography{references}

\end{document}